\documentclass{article}

\usepackage{amssymb,rotating,harvard}

\citationstyle{dcu}

\author{Sergej V. Aksenov \and Michael A. Savageau\thanks{Corresponding author}}

\title{Statistical inference and modeling with the {\em S}~distribution}

\date{Department of Microbiology and Immunology, The University of Michigan, Ann Arbor, MI 48109\thanks{Email: aksenov@umich.edu (Sergej V. Aksenov), savageau@umich.edu (Michael A. Savageau)}}

\begin{document}

\maketitle

\thispagestyle{myheadings}
\markboth{}{Submitted to J. Statist. Comput. Simul.}

\abstract{We consider the problem of statistical inference for the {\em S} distribution and introduce new minimum distance estimators for the four parameters of the {\em S} distribution using Kolmogorov-Smirnov, Cram\'{e}r-von Mises and related distance metrics.  Approximate goodness-of-fit and confidence intervals for parameters are calculated using bootstrap methods.  We discuss further how the {\em S} distribution can be used to solve various problems of statistical modeling associated with parameter inference, including goodness-of-fit tests, Monte Carlo simulations and modeling trends in the distributions.}

{\bf Keywords}: {\em S} distribution, minimum distance estimators, goodness-of-fit, bootstrap

\section{Introduction}

Parameter estimation is important for solving various problems associated with statistical inference, e.g. for hypothesis testing and for Monte Carlo and stochastic modeling.  In practical terms, the very first step in such modeling involves choosing a distribution function (d.f.) for the probability law that best describes the random process and/or available data.  Very often one is presented with a poorly understood process to model and with samples of data of realistic (moderate) size, and then this choice can be far from unique.  In other words, several distributions in common use (e.g., Normal, Logistic, Weibull, Laplace etc.) often can be used with almost equal success.  This difficulty can be alleviated by using distributional families.  Such families however tend to be unwieldy mathematically and even present serious compuational difficulties, for example, poles in the Pearson system of distributions.

A useful approach for these problems has been developed over the past ten years.  It involves a univariate continuous four-parameter distributional family called the {\em S} distribution~\cite{savageau82} that is capable not only of {\em approximating} many central and noncentral unimodal univariate distributions rather well~\cite{voit92}, but also of representing an uncountable multitude of others as its parameters change smoothly~\cite{sorribas00}.   It includes Exponential, Logistic, Uniform and Linear distributions as special parametric cases.  The {\em S} distribution derives its name from the fact that it is based on the theory of S-systems~\cite{savageau76,voit91}.  The versatility and relative mathematical simplicity of the {\em S} distribution prompts for its use in statistical inference problems.  We note here that the errors resulting from approximation seem to be a small price relative to the advantage of having a ``best fit'' distribution readily available for a particular problem.  A number of parameter estimation techniques have been proposed over the years (reviewed briefly below).

In this article we propose minimum distance (MD) estimators for the {\em S} distribution parameters that make use of goodness-of-fit statistics of supremum (Kolmogorov-Smirnov and Kuiper) and quadratic (Cram\'{e}r-von Mises and Watson) types as distance metrics between empirical d.f. and {\em S} distribution d.f. defined by Equation~(\ref{df}) below.  Note that there is no categorization of data involved.  The MD estimators then can be used in testing the goodness-of-fit hypothesis.  We propose bootstrapping to approximate critical values for a goodness-of-fit test and then calculate approximate confidence intervals for the parameter estimates.  Neither goodness-of-fit nor accuracy of estimates have previously been evaluated for {\em S} distribution estimations.  We illustrate the new method with examples of parameter inference using data generated from the {\em S} distribution.  We conclude with a discussion of the uses of the {\em S} distribution in statistical modeling problems.  A computational realization of the proposed estimation procedure is described elsewhere~\cite{aksenov01}.  

\section{The {\em S} distribution}

The {\em S} distribution is defined in terms of its d.f. $F(x)$~\cite{savageau82}, which is the solution of the following initial value problem (i.v.p.) for an ordinary differential equation (o.d.e.)
\begin{equation} 
\label{df}
f(x) = \frac{dF}{dx} = \alpha \left( F^g - F^h \right), \qquad F(x_0)=F_0
\end{equation}
Note that the right-hand side of Equation~(\ref{df}) is also the probability function (p.f.) $f(x)$, which is thus an algebraic function of the d.f.  The {\em S} distribution has four parameters 
\begin{equation}
\label{theta}
\theta=(g,h,\alpha,x_0)
\end{equation}
with $x_0$ for location, $\alpha$ for scale and $g$ and $h$ for shape.  Often it is convenient to choose $x_0$ as a median, $F_0=0.5$.

The {\em S} distribution in Equation~(\ref{df}) is completely specified by its four parameters $\theta$.  Conditions on the parameters, $\alpha > 0$ and $g<h$, ensure that $F(x)$ is a proper d.f., i.e. a monotone function of the random variable $x$ with $F(-\infty)=0$ and $F(\infty)=1$.

We can obtain a simple condition on the parameters that provides for unimodality of the {\em S} distribution.  Differentiating the p.f. in~(\ref{df}) for $x$ and equating to zero we obtain
\begin{equation}
\label{mode}
\frac{g}{h} = F^{h-g}
\end{equation}
One can immediately see that, given $g$ and $h$, this equation has a real solution $F$ (and hence mode $x_m$) if $g$ and $h$ are both positive, $g>0$ and $h>0$.  Otherwise, the {\em S} distribution has a half-mode (i.e., is J-shaped).  This follows from the left-hand side of Equation~(\ref{mode}) being negative if $g$ and $h$ have different signs, and greater than one if $g$ and $h$ are both negative.  The right-hand side of Equation~(\ref{mode}) has a positive real value between 0 and 1 for any $g<h$.

We can define skewness of the {\em S} distribution by evaluating the d.f. at the mode~\cite{savageau82}: 
\begin{equation} 
\label{fmode}
S=F(x_m)=\left( \frac{g}{h} \right)^{1/(h-g)}
\end{equation}
Now the {\em S} distribution is skewed to the right if the mode is less than the median, $S<0.5$, is symmetrical if the mode coincides with the median, $S=0.5$, and is skewed to the left if the mode is greater than the median, $S>0.5$.  For negative $g$, the {\em S} distribution is skewed to the right.

Moments of the {\em S} distribution can be obtained numerically by integrating the p.f. 
\begin{equation} 
\label{mom}
\mu_r^{\prime} = E(x^r) = \alpha \int_{-\infty}^{\infty} x^r \left( F^g(x) - F^h(x) \right) dx
\end{equation}
simultaneously with the solution of the o.d.e.~(\ref{df}).

Quantiles of the {\em S} distribution can be obtained by using the fact that there is a monotone one-to-one relation between the random variable and the d.f.  Thus, we can rewrite the o.d.e.~(\ref{df}) as
\begin{equation} 
\label{quantile}
\frac{dx}{dF} = \frac1{\alpha} \frac1{F^g - F^h}, \qquad x(F_0)=x_0
\end{equation}
The solution of this equation can be obtained numerically.  A closed form solution can be found in terms of elementary transcendental functions for a certain subclass of parameters $g$ and $h$~\cite{voit84} or in terms of Lerch's transcendent for all $g,h\in \mathbb{R}$~\cite{hernandez01}.  This is done by separating variables in either o.d.e.~(\ref{df}) or~(\ref{quantile}) and integrating to obtain the following 
\begin{equation} 
\label{sepint}
x(F) = x_0 + \frac1{\alpha} \int_{F_0}^{F} \frac{dt}{t^g-t^h}
\end{equation}
\citeasnoun{voit84} solved the integral in Equation~(\ref{sepint}) in terms of elementary transcendental functions for $g,h\in \mathbb{R}$, when $g=(h \sigma -1)/(\sigma-1)$ and a signed rational number $\sigma \neq 1$, or for $g,h\in \mathbb{R}$, when $g<h=1$ and $\sigma = 1$.

\citeasnoun{hernandez01} represented the integrand in Equation~(\ref{sepint}) as an infinite sum to obtain
\begin{equation} 
\label{qsol}
x(F) = x_0 + \frac1{\alpha} \sum_{k=0}^{\infty} \int_{F_0}^{F} t^{k (h-g)-g} dt
\end{equation}
The sum converges to a finite value if $k (h-g)-g \neq -1$, which covers $g,h\in \mathbb{R}$ when $g<h \leq 1$ or $1<g<h$ such that $g \neq (h k+1)/(k+1)$ for $k\in \mathbb{N}$.  This condition defines a ``generic'' quantile solution.  In the ``generic'' case, the quantile function is expressed in terms of Lerch's transcendent as follows:
\begin{eqnarray} 
\label{qgen}
x & = & x_0 + \frac1{\alpha (1-g)} \left( F^{1-g} \Phi\left(F^{h-g},1,\frac{1-g}{h-g} \right) - \right. \nonumber \\
&& \left. F_0^{1-g} \Phi\left(F_0^{h-g},1,\frac{1-g}{h-g} \right) \right)
\end{eqnarray}
where Lerch's transcendent $\Phi(z,s,v)$ is defined by the following series~\cite{magnus66}:
\begin{equation} 
\label{lerchphi}
\Phi(z,s,v) = \sum_{k=0}^{\infty} \frac{z^k}{(v+k)^s},\qquad |z|<1,\qquad v \neq 0,-1,\dots
\end{equation}
In the ``nongeneric'' case, which covers $g,h\in \mathbb{R}$ when $1\leq g<h$ and $g=(h k+1)/(k+1)$ for $k\in \mathbb{N}$, the integral in~(\ref{qsol}) produces a logarithmic term that has to be integrated separately.  For $g=1$, the logarithmic term is at $k^{\ast}=0$ and so the quantile function is
\begin{equation} 
\label{qnongen1}
x = x_0 + \frac1{\alpha} \left( \log \frac{F}{F_0} + \frac{\left(1-F_0^{h-1}\right)/\left(1-F^{h-1}\right)}{h-1} \right)
\end{equation}
and for $g>1$, the logarithmic term is at $k^{\ast}=(g-1)/(h-g)$ and the quantile function is
\begin{equation} 
\label{qnongen2}
x = x_0 + \frac1{\alpha} \left( \log \frac{F}{F_0} + \sum_{k=0,k\neq k^{\ast}}^{\infty} \frac{F^{k (h-g)-g+1} - F_0^{k (h-g)-g+1}}{k (h-g)-g+1} \right)
\end{equation}
Note that the ``nongeneric'' solution is identical to the one found by \citeasnoun{voit84} for $\sigma = -k$.

The explicit quantile function given by Equations~(\ref{qgen}), (\ref{qnongen1}) and~(\ref{qnongen2}) has several important consequences for both parameter inference and modeling.  First, it can be shown that $\lim_{F\rightarrow 0+} x(F)= constant$ for $g<1$ (the ``generic'' case).  In other words, there is no infinite tail for the corresponding {\em S} distribution, which then becomes left-truncated.  The truncation point can be calculated by letting $F=0$ in Equation~(\ref{qgen})
\begin{equation} 
\label{q0}
x^{\ast} = x_0 - \frac{F_0^{1-g} }{\alpha (1-g)} \Phi\left(F_0^{h-g},1,\frac{1-g}{h-g} \right)
\end{equation}
Second, the existence of a finite $x^{\ast}$ calls for care when solving o.d.e.~(\ref{df}) numerically, because the solution of the o.d.e. is not unique at $x^{\ast}$ where the solution $F(x^{\ast})=0$ joins the trivial solution $F=0$.  Nonuniqueness can be formally checked by showing that the Lipschitz condition is not satisfied at $x^{\ast}$.   Most o.d.e. solvers, whose algorithms assume uniqueness of the solution, will have trouble converging near $x^{\ast}$.  Third, an explicit expression for the mode of the {\em S} distribution, if it exists, is now possible by substituting~(\ref{fmode}) into Equations~(\ref{qgen}), (\ref{qnongen1}) and~(\ref{qnongen2}):
\begin{eqnarray} 
\label{explmode}
x_m & = & x_0 + \frac1{\alpha (1-g)} \left( \left(\frac{g}{h}\right)^{(1-g)/(h-g)} \Phi\left(\frac{g}{h},1,\frac{1-g}{h-g} \right) - \right. \nonumber \\
&& \left. F_0^{1-g} \Phi\left(F_0^{h-g},1,\frac{1-g}{h-g} \right) \right) \nonumber \\
x_m & = & x_0 + \frac1{\alpha} \left( \log \frac1{F_0} \left(\frac{g}{h}\right)^{1/(h-g)} + \right. \nonumber \\
&& \left. \frac{\left(1-F_0^{h-1}\right)/\left(1-\left(\frac{g}{h}\right)^{(h-1)/(h-g)}\right)}{h-1} \right) \nonumber \\
x_m & = & x_0 + \frac1{\alpha} \left( \log \frac1{F_0} \left(\frac{g}{h}\right)^{1/(h-g)} + \right. \nonumber \\ 
&& \left. \sum_{k=0,k\neq k^{\ast}}^{\infty} \left( \left(\frac{g}{h}\right)^{(k (h-g)-g+1)/(h-g)} - F_0^{k (h-g)-g+1}\right) \times \right. \nonumber \\
&& \left. \frac1{k (h-g)-g+1} \right)
\end{eqnarray}
Finally, the quantile function makes it easy to use a direct inversion method for the generation of random variates from the {\em S} distribution.

\section{Methods for parameter inference}

Existing techniques for estimating parameters of the {\em S} distribution, given a random sample $\{x_i\}$ $i=1,\dots,n$, have been based on graphical, nonlinear regression~\cite{voit92}, and maximum likelihood (ML)~\cite{voit00a} methods.  The graphical method is relatively straightforward.  As $F\rightarrow 0+$ the term $F^g$ dominates over the term $F^h$ and the plot of $\ln f$ vs. $\ln F$ is a straight line $\ln f = \ln \alpha + g \ln F$ with slope $g$ and intercept $\alpha$.  Then, given $F$ at the inflection point, which corresponds to the mode (if it exists), and the previously estimated $g$, one can estimate $h$ from Equation~(\ref{mode}).

Nonlinear regression can be accomplished with $x$ vs. $F$ of the o.d.e. or with an equivalent representation in terms of the algebraic equation $f$ vs. $F$.  Regression of the algebraic equation seems to be faster and less numerically involved: however, estimating $x_0$ is not possible.  During regression, the residual squared error (r.s.e.) is typically minimized and data is represented in a categorical form.  In a recent example of regression-based estimation, \citeasnoun{sorribas00} propose to estimate $x_0$ by the sample median, to fix one of the parameters (e.g., $\alpha$ as the inverse of the sample standard deviation), and to fit the remaining pair (e.g., $g$ and $h$) using the o.d.e.~(\ref{df}) and suitably categorized data.  This procedure helps to avoid the algorithmic difficulties associated with fitting four parameters simultaneously.  Fixing some parameters and fitting the others exposed correlations between the ``best-fit'' parameters.  For example, for a given random sample, fixing $\alpha$ at increasing values produced pairs of $g$ and $h$ where $g$ was increasing and $h$ was decreasing.  In addition, one observed uncertainties in the ``best-fit'' parameters that resulted from optimization runs being initialized with different values, and from sampling variability.

Drawing on the closed-form quantile function given by Equations~(\ref{qgen}), (\ref{qnongen1}) and~(\ref{qnongen2}), \citeasnoun{hernandez01} proposed to use least-squares fitting of the {\em S} distribution quantiles to the sample quantiles.  Theoretical {\em S} distribution quantiles are evaluated at values of the empirical d.f., which is a step function with jumps at the data points.

Recently, \citeasnoun{voit00a} introduced a ML procedure to calculate estimates for $g$ and $h$.  After using the algebraic relationship between the p.f. and the d.f., the log-likelihood function is
\begin{equation} 
\label{loglike}
\log L(\theta) = n \log \alpha + \sum_{i=1}^n \log \left( F^g(x_i;\theta) - F^h(x_i;\theta) \right)
\end{equation}
Direct minimization of the function~(\ref{loglike}) involves  numerical solution of the o.d.e.~(\ref{df}) (unpublished computational realization by Voit and Schwacke; also by Sorribas, personal communication).  However, Voit suggested an approximate ML method that requires only solution of nonlinear algebraic equations.  First, he replaces the theoretical {\em S} distribution d.f. evaluated at data points $F(x_i;\theta)$ by the empirical d.f. $\hat{F}$ that is its consistent estimate.  Second, he introduces a constraint on parameters in the form of a fixed integral in the phase space $dF/dx$ vs. $F$, which makes $\log L(\theta)$ a function of only $g$ and $h$.  Differentiating $\log L$ with respect to $g$ and $h$ and equating the derivatives to zero results in nonlinear equations for $g$ and $h$ that are then solved iteratively:
\begin{eqnarray} 
\label{ghmle}
0 & = & \frac1{g+1} + \frac1{h+1} + \frac{\sum_{i=1}^{n} \log \hat{F}(x_i)}{n} \nonumber \\
0 & = & \frac1{h+1} - \frac1{h-g} + \frac{\sum_{i=1}^{n} \log \hat{F}(x_i) / (1-\hat{F}^{g-h}(x_i))}{n}
\end{eqnarray}
Note that the last point of the ordered sample causes a discontinuity because $\hat{F}(x_n)=1$ by definition.  Equations~(\ref{ghmle}) can still be solved numerically, provided one uses L'Hospital's rule to evaluate the last term of the sum:
\begin{equation}
\label{lhospital}
\lim_{y\rightarrow 1-} \frac{\log y}{1-y^{g-h}} = \lim_{y\rightarrow 1-} \frac{1/y}{(h-g) y^{g-h-1}} = \frac1{h-g}
\end{equation}

In summary, existing techniques for {\em S} distribution parameter estimation tend to use categorized data and generally lack goodness-of-fit information.  These deficiencies motivated us to develop MD estimators.

\section{Minimum distance estimators}

The MD method was introduced by \citeasnoun{wolfowitz57} and since then has proved to be a convenient method for strongly consistent parameter estimation.  The idea of the method is to match the empirical d.f. to a theoretical one as closely as possible, using a distance function $\delta(\cdot,\cdot)$.  In the context of the {\em S} distribution we have the d.f.~(\ref{df}) defined on $\theta \in \Theta$ where $\Theta$ is the following subset of $\mathbb{R}^4$:
\begin{equation} 
\label{thetadom}
\Theta := \{\theta: \alpha>0,g<h,h\in \mathbb{R}, x_0 \in \mathbb{R},C(\theta,x_i)\leq 0\}
\end{equation}
where the nonlinear constraint function $C$ is defined as
\begin{equation} 
\label{constr}
C = x_0 - \mathrm{min}_{i=1,\dots,n}(x_i) + \frac{F_0^{1-g} }{\alpha (1-g)} \Phi\left(F_0^{h-g},1,\frac{1-g}{h-g} \right)
\end{equation}
The nonlinear constraint~(\ref{constr}) is essential for estimation because it ensures that the {\em S} distribution is consistent with the data at all times, and at the optimum vector $\hat{\theta}$ in particular.  This means that the truncation point, which is finite, is less than the minimum observed data point and thus that the d.f. is defined at all data points.  Of course this does not ensure against the possibility that an even lower data point just has not been observed and thus that the true population should be truncated at an even lower value of $x$, or not truncated at all.  Note that evaluation of the constraint function $C$ depends on an accurate and fast method for calculation of Lerch's transcendent $\Phi(z,s,v)$. This is now possible with recent advances in convergence acceleration techniques~\cite{aksenov01,jentschura99,jentschura01}.  The d.f. is calculated at $x_i$ using Equation~(\ref{df}) as long as the desired solution is not too close to the truncation point $x^{\ast}$.  In the immediate proximity of $x^{\ast}$ one can solve the nonlinear (transcendental) equation~(\ref{qgen}) for $F$.

The empirical d.f. is defined as a step function
\begin{equation} 
\label{edf}
\hat{F}(x) = \frac{\#(x_i\leq x)}{n}
\end{equation}
where $\#(\cdot)$ signifies the number of $x_i$ less than or equal to $x$ and weight $1/n$ is put on each point; if there are tied observations, proportionately more weight is put on the unique points.

Given the distance metric $\delta (\hat{F},F)$, a MD estimator $\hat{\theta}$ is given by the solution of the following equation
\begin{equation} 
\label{disteq}
\delta(\hat{F}(x_i),F(x_i;\hat{\theta})) = \mathrm{inf}_{\theta \in \Theta} \delta(\hat{F}(x_i),F(x_i;\theta))
\end{equation}

As the distance metric $\delta$, we consider here four goodness-of-fit statistics of the supremum (Kolmogorov-Smirnov and Kuiper) and quadratic (Cram\'{e}r-von Mises and Watson) types~\cite{agostino86}.  This allows us to combine estimation with testing the validity of fit.  The Kolmogorov-Smirnov statistic $D$ is the largest unsigned vertical distance between $\hat{F}$ and $F$, the Kuiper statistic $V$ is the sum of the largest signed vertical distances, the Cram\'{e}r-von Mises statistic $W^2$ is the integral of the squared differences between $\hat{F}$ and $F$, and the Watson statistic $U^2$ is a modified version of $W^2$:
\begin{eqnarray} 
\label{genstat}
D & = & \mathrm{sup}_{x} |\hat{F}(x) - F(x)| \nonumber \\
V & = & \mathrm{sup}_{x} \left( F(x) - \hat{F}(x) \right) + sup_{x} \left( \hat{F}(x) - F(x) \right) \nonumber \\
W^2 & = & n \int_{-\infty}^{\infty} \left(  \hat{F}(x) - F(x) \right)^2 dF(x) \nonumber \\
U^2 & = & n \int_{-\infty}^{\infty} \left(  \hat{F}(x) - F(x) - \right. \nonumber \\
&& \left. \int_{-\infty}^{\infty} \left(  \hat{F}(x) - F(x) \right) dF(x)\right)^2 dF(x)
\end{eqnarray}
For the one-sample problem that we are dealing with, computational formulas for the statistics can be derived from~(\ref{genstat}) using the probability integral transformation $z = F(x)$, where $z$ is uniformly distributed, and letting $z_i = F(x_i)$:
\begin{eqnarray} 
\label{compstat}
D & = & \mathrm{max} \left( \mathrm{max}_{i=1,\dots,n} \left(\frac{i}{n} - z_i \right), \mathrm{max}_{i=1,\dots,n} \left( z_i - \frac{i-1}{n} \right) \right) \nonumber \\
V & = & \mathrm{max}_{i=1,\dots,n} \left(\frac{i}{n} - z_i \right) + \mathrm{max}_{i=1,\dots,n} \left( z_i - \frac{i-1}{n} \right) \nonumber \\
W^2 & = & \frac1{12 n} + \sum_{i=1}^n \left( z_i - \frac{2 i-1}{2 n} \right)^2 \nonumber \\
U^2 & = & W^2 - n \left( \sum_{i=1}^n \frac{z_i}{n} - 0.5 \right)^2
\end{eqnarray}

The MD estimators $\hat{\theta}$ obtained as a solution of Equations~(\ref{thetadom}), (\ref{constr}) and (\ref{disteq}) and with metrics $\hat{\delta}$~(\ref{compstat}) can be used in testing the goodness-of-fit.  Formally, we wish to test the composite hypothesis that the random sample $\{x_i\}$ comes from the {\em S} distribution:
\begin{equation} 
\label{hyp}
H_0:F \in \mathcal{F}
\end{equation}
against general alternatives, where $\mathcal{F}$ is the class of the {\em S} distribution d.f.s
\begin{equation}
\label{fclass}
\mathcal{F} := \{F(\cdot,\theta):\theta \in \Theta\}
\end{equation}

Now given parameter estimates $\hat{\theta}$, one calculates the empirical d.f.-based goodness-of-fit statistic $\hat{\delta}$~(\ref{compstat}) and compares it with the critical point corresponding to a specified significance level, typically 0.01 or 0.05.  For the so-called case 0, when the distribution $F(x)$ is completely specified, asymptotic distributions of goodness-of-fit statistics are known and critical points have been tabulated~\cite{stephens70,stephens74}.  However, in the general case when parameters are estimated from data, distributions of statistics have to be approximated.  A relatively straightforward though computationally-intensive way of obtaining the critical points is to approximate sampling distributions of goodness-of-fit statistics by the bootstrap method.  The asymptotic validity of the bootstrap method for MD goodness-of-fit tests was established in~\cite{beran86} and for more general problems in~\cite{romano88}.  With the bootstrap method, one can also calculate approximate confidence intervals for parameter estimates.   The MD estimators have been shown to have an asymptotic distribution~\cite{sahler70,bolthausen77}.

The bootstrapping algorithm is as outlined in~\cite{efron93}.  One samples $B$ times with replacement, either from an empirical d.f.~(\ref{edf}) of the sample (in a nonparametric mode) or from the parametric model with parameters $\hat{\theta}$ (in a parametric mode), and calculates parameter estimates and goodness-of-fit statistics exactly the same way as with the original sample.  These are now called bootstrap replications $\hat{\theta}^{\ast}$ and $\hat{\delta}^{\ast}$.  The lower and upper critical values for a goodness-of-fit statistic $\delta$, corresponding to a significance level $\alpha<0.5$, are then the $k=[(B+1) \alpha/2]$th and  $(B+1-k)$th largest values of the ordered bootstrap replications $\hat{\delta}^{\ast}$, respectively, where $[\cdot]$ signifies taking the integer part.  The observed statistic value $\hat{\delta}$ is then compared with the critical values.  Comparison with the lower critical value ensures against the so called superuniformity when the statistic takes too small a value~\cite{stephens70}.  Equivalently, one can calculate the achieved significance level (a.s.l.) of a statistic $\hat{\theta}$ that is simply the empirical quantile based on an ordered sample of replications $\hat{\delta}^{\ast}$.  The a.s.l. is then compared with the specified significance level of the test.

To obtain a $(1-\alpha )100\%$ equitailed bootstrap-percentile confidence interval for the parameter estimates, bootstrap replications $\hat{\theta}^{\ast}$ are ordered, and lower $\theta_{\mathrm{lo}}^{\ast}$ and upper $\theta_{\mathrm{up}}^{\ast}$ endpoints of the interval are again estimated by the $k=[(B+1) \alpha /2]$th and $(B+1-k)$th largest values.  However, bootstrap-percentile intervals can have substantial coverage error as shown in the following equation
\begin{equation} 
\label{coverr}
P(\theta_{\mathrm{lo}} < \theta < \theta_{\mathrm{up}}) = 1-\alpha + O(f(n))
\end{equation}
The bootstrap-percentile method is first-order accurate in that the rate with which the coverage error goes to zero is $f(n)=n^{-1/2}$, as the sample size goes to infinity.   To improve the accuracy, the BCa (bias-corrected and accelerated) method was proposed~\cite{efron87}.  In this method, the endpoints $\theta_{\mathrm{lo}}^{\ast}$ and $\theta_{\mathrm{up}}^{\ast}$ are calculated as empirical $\alpha_1$th and $\alpha_2$th quantiles, respectively, 
\begin{eqnarray} 
\label{bca}
\alpha_1 & = & \Phi\left( \hat{z}_0+\frac{\hat{z}_0+z^{\alpha /2}}{1-\hat{a} (\hat{z}_0+z^{\alpha /2})} \right) \nonumber \\
\alpha_2 & = & \Phi\left( \hat{z}_0+\frac{\hat{z}_0+z^{1-\alpha /2}}{1-\hat{a} (\hat{z}_0+z^{1-\alpha /2})} \right)
\end{eqnarray}
where $\Phi$ is the standard Normal d.f. and $z^{\alpha}$ is the $\alpha$th quantile of the standard Normal distribution, i.e. $\Phi(z^{\alpha})=\alpha$.  The bias-correction constant $\hat{z}_0$ is obtained as the proportion of bootstrapped replications that are less than the observed value,
\begin{equation} 
\label{biasz0}
\hat{z}_0 = \Phi^{-1} \left( \frac{\#\{\hat{\theta}^{\ast}<\hat{\theta}\}}{B} \right)
\end{equation}
where $\Phi^{-1}$ is the Normal quantile function (i.e., the inverse of the d.f.).  The acceleration connstant $\hat{a}$ can be estimated in terms of the jackknife values $\hat{\theta}_{(i)}$ ($\theta$s estimated from the sample omitting the $i$th point):
\begin{equation} 
\label{acc}
\hat{a} = \frac{\sum_{i=1}^n \left( \sum_{i=1}^n \hat{\theta}_{(i)}/n - \hat{\theta}_{(i)} \right)^3}{6 \left( \sum_{i=1}^n \left( \sum_{i=1}^n \hat{\theta}_{(i)}/n - \hat{\theta}_{(i)} \right)^2 \right)^{3/2}}
\end{equation}
The BCa intervals are second-order accurate in that the coverage error goes to zero with rate $f(n)=n^{-1}$. 

Like all bootstrap estimates, the confidence interval endpoints have variance that is due to sampling error and bootstrap resampling error.  We can estimate the variance of endpoints (which are sample quantiles) using the jackknife-after-bootstrap method in which for each $i$th data point from the ordered sample, one groups bootstrap resamples that do not contain that particular point and calculates confidence interval endpoints exactly as above over that collection of resamples, $\hat{\theta}_{B(i)}$~\cite{efron92}.  The estimate of variance is then
\begin{equation} 
\label{bootvar}
\mathrm{var}(\hat{\theta}) = \frac{n-1}{n} \sum_{i=1}^n \left( \hat{\theta}_{B(i)} - \frac{\sum_{i=1}^n \hat{\theta}_{B(i)}}{n} \right)^2
\end{equation}
One can use Equation~(\ref{bootvar}) to decide if a given number of resamples $B$ is satisfactory by calculating the coefficient of variation, $\mathrm{cv}=\mathrm{var}^{1/2}(\hat{\theta})/\hat{\theta}$, as a function of $B$ and choosing a threshold for $\mathrm{cv}$, say 0.1 (which means we are unwilling to accept more than 10\% of contribution of Monte Carlo error to the estimate).

The BCa intervals can be quite expensive to calculate, especially taking into account the optimization step in Equations~(\ref{thetadom}), (\ref{constr}) and (\ref{disteq}).  In general, on the order of 1000 resamples might be commonly needed to achieve a small Monte Carlo error~\cite{efron87}.  

However, one can focus on coverage accuracy of the bootstrap approximation and, instead of accumulating more bootstrap resamples to reduce the error, use the number of resamples as a calibration parameter to achieve a specified coverage.  Such an approach leads to the extreme bootstrap percetiles method~\cite{lee00}.  As a first step in constructing the equitailed percentile interval of nominal coverage $(1-\alpha)100\%$, one solves the following equations for the minimum required number of resamples $B$
\begin{eqnarray} 
\label{extremeeq}
\alpha/2 & = & \frac1{B+1} + \frac{\hat{a} b^3}{B} \nonumber \\
\alpha/2 & = & \frac1{B+1} - \frac{\hat{a} b^3}{B}
\end{eqnarray}
where $b$ is the positive solution of equation
\begin{equation} 
\label{bb}
B \phi(b-b^{-1})=b
\end{equation}
and $\phi$ is the standard Normal p.f.  The maximum of the two solutions for Equations~(\ref{extremeeq}) and~(\ref{bb}) is then the derived minimum number of resamples.  Equations~(\ref{extremeeq}) and~(\ref{bb}) are obtained from asymptotic expansions of the extreme coverage associated with the bootstrap-percentile method, and assume the validity of Edgeworth expansions for the bootstrap distributions of the standardized bootstrapped statistic and the smooth model for the statistic as a function of the mean.  The validity of these equations is thought however to extend to more general statistical functionals~\cite{lee00} and thus they are likely to be applicable here.  The coverage error of the extreme percentiles intervals goes to zero with rate $f(n)=n^{-1/2} \log^{1/2} n$, which is slower than with the BCa intervals.  the extreme percentile intervals can however provide a large reduction of computational effort because the minimum required $B$ will typically be much less than 1000.  As with all bootstrap estimates, it is worth looking at histograms of replications; highly skewed histograms are indicative of inaccurate estimates of the tails of the bootstrap distributions and of the requirement for more simulation effort.  Also note that the confidence intervals here are the marginal intervals for  the parameters $\theta$, constructed from a multivariate empirical bootstrap distribution.  Construction of the simultaneous confidence regions would be rather awkward given the dimension of $\theta$.  Confidence regions for two-parameter location and scale families based on the Kolmogorov-Smirnov statistic have been considered in~\cite{easterling76,littel78}.

\section{Example: inference from the {\em S} distribution data}

Here we apply the MD estimators, derived using the above procedure, to parameter inference for a random sample generated from a specified {\em S} distribution.  We use a computational realization of this procedure that is a collection of Mathematica and C programs~\cite{aksenov01}.  A Mathematica notebook documenting all the calculation in this section is available from the corresponding author.

We generate a random sample of size $n=100$ from an {\em S} distribution with the parameters in~(\ref{theta}) given by $\theta=(0.5,1.6,1.0,0.0)$. For the particular random sample used here, the seed for the Mathematica random number generator was 11235.  This distribution is left-truncated with truncation point $x^{\ast}=-1.70745$ (see Equation~(\ref{q0})) and unimodal with the mode at $x_m=-0.38601$ (see Equation~(\ref{explmode})).

As a first attempt we estimate parameters using a combination of existing methods: let $\hat{x}_0$ be the sample median, $\hat{\alpha}$ be the inverse of the standard deviation of the data and $\hat{g}$ and $\hat{h}$ be the approximate maximum likelihood estimates calculated using Equations~(\ref{ghmle}).  These estimates and the four goodness-of-fit statistics~(\ref{compstat}) are shown in the third column of Table~\ref{pointexample}.  We do not adjust $\hat{\alpha}$ and $\hat{x}_0$ to have better agreement with the data as advised in~\cite{voit00a} since these estimates serve only as initial guesses for the optimization.  These initial estimates are reasonably close to the population parameters for this particular sample, but of course we are more interested in how the estimators behave in the long run when applied to other samples from the same population.  The goodness-of-fit statistics calculated with these estimates are much larger than those calculated with the true parameters, but again we do not know how reproducible this diffrence is in the long run.

The MD estimates obtained by using Equations~(\ref{thetadom}), (\ref{constr}) and (\ref{disteq}) with each of the four distance metrics~(\ref{compstat}) are shown in the last four columns of Table~\ref{pointexample}.  The values of the goodness-of-fit statistics are substantially lower than those calculated with the first estimates or with the true parameters.  We are now ready for evaluating the goodness-of-fit with the bootstrap method.

The equitailed extreme-percentile confidence intervals with intended coverage of 95\% for MD estimators with the four goodness-of-fit statistics are shown in Table~\ref{extremeexample}.  According to Equations~(\ref{extremeeq}) and~(\ref{bb}) with $\alpha=0.05$, only 39 bootstrap resamples were needed for the approximation.  We performed calculations with nonparametric resampling from the empirical d.f. and parametric resampling from the {\em S} distribution d.f. with MD estimates $\hat{\theta}$ obtained using the corresponding goodness-of-fit statistics.  Both nonparametric and parametric intervals have similar lengths and shapes (data not shown).  Note that the MD estimates based on the different goodness-of-fit statistics have intervals of different lengths.  For example, the quadratic statistics $W^2$ and $U^2$ give intervals for $h$ and $\alpha$ that are wider than those provided by the supremum statistics $D$ and $V$.  Also, all estimators have the true values inside the intervals, except for the parametric interval with Kuiper for $\alpha$, which indicates a possible bias for this estimator.  Observed values of the goodness-of-fit statistics are within their respective 95\% intervals, indicating that the null hypothesis~(\ref{hyp}) cannot be rejected at the 0.05 significance level.  We note that the a.s.l.s for nonparametric bootstraping with supremum statistics are somewhat lower than those for parametric bootstrapping, making the nonparametric tests conservative.  This observation is reversed for the quadratic statistics.

An alternative way to calculate approximate confidence intervals is with the  BCa method.  Table~\ref{bcaexample} shows 95\% equitailed intervals obtained with 4000 parametric and nonparametric bootstrap resamples by using Equations~(\ref{bca}), (\ref{biasz0}) and~(\ref{acc}).  Note again that the supremum statistics are more conservative for nonparametric than for parametric tests.  The availablility of 100 times more resamples than with the extreme-percentile method permits more thorough investigation.  For example, bootstrap distributions of estimates differ radically for the two types of functionals (data not shown).  For supremum functions $D$ and $V$, distributions of all estimates are more or less symmetric with moderate tails.  In contrast, for quadratic functions $W^2$ and $U^2$, distributions are highly skewed.  This leads to wide intervals for $\hat{h}$ and $\hat{\alpha}$.  Curiously, distributions for $\hat{g}$ are bimodal, indicating the presence of at least two local minima in the optimization problem involving quadratic functions.  Distributions for $\hat{x}_0$ are nearly symmetrical in all cases.  High skeweness of distributions is accompanied by high variability of bootstrap estimates for the endpoints of the confidence intervals.  The overall variability is expected to settle down at the level of sampling variability with increasing number of resamples $B$.  This indeed happens for the upper endpoints of the estimation based on the  Kolmogorov-Smirnov distance function, but not for any of the lower endpoints, and the pattern is more erratic as we move to the estimates based on the quadratic functions (data not shown).

\begin{figure}[t]
\includegraphics[width=0.9\textwidth]{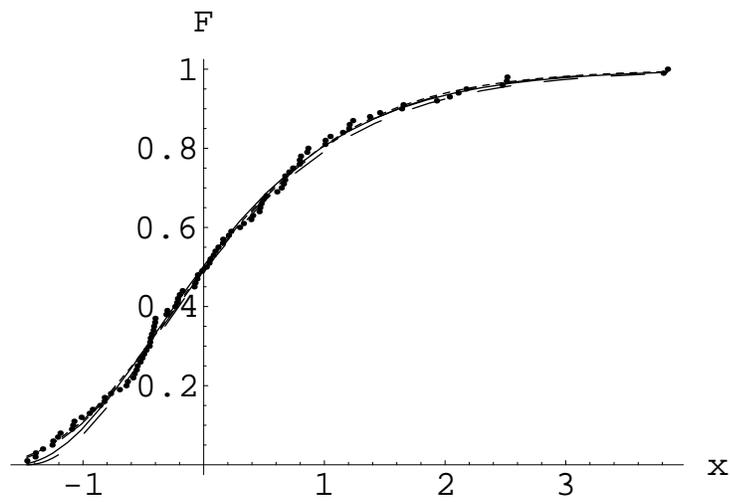}
\caption{Empirical (dots) and theoretical {\em S} distribution (lines) d.f.s calculated with the MD estimates.  See Table~\ref{pointexample} for population and estimated parameters.} \label{fig1}
\end{figure}

Plots of the empirical d.f. and the ``best-fit'' {\em S} distribution d.f. are shown in Figure~\ref{fig1}.  While all four MD estimators give visually good approximations, which is also evident from the close agreement among the estimates and the population values, the properties of the estimators are strikingly different as discussed above.

\section{Existing applications of the {\em S} distribution in statistical modeling and suggested extensions}

The ability of the {\em S} distribution to approximate diverse distributional forms suggests its use in various stochastic models that give rise to univariate unimodal distributions.  Methods of parameter inference for the {\em S} distribution, including the MD estimators proposed in this article, are of course critical to any data-based modeling of this sort.

One of the main application areas is Monte Carlo modeling.  Specifically, one might be interested in repeated sampling from an {\em S} distribuion that is the best numerical model for random data.  Apart from parameter estimation, efficient random number generation from an {\em S} distribution is required.  An example of {\em S} distribution modeling in risk assessment studies is provided by \citeasnoun{voit00c}.  They also described an approximate method to sample from an {\em S} distribution, by interpolating among tabulated {\em S} quantiles with a rational function.  An exact method is to use inversion with the quantile functions given by Equations~(\ref{qgen}), (\ref{qnongen1}) and~(\ref{qnongen2})~\cite{hernandez01}.  In risk analysis applications, the parameters of risk models are often uncertain and it is desirable to investigate the sensitivity of risks to these parameters.  A fundamental approach is to assign an {\em S} distribution to parameters of the risk model and simulate it many times in an attempt to evaluate the statistcal uncertainties in the risk as a function of the input distribution of the parameters.  A similar application of {\em S} distributions can be found for hierarchical Monte Carlo simulations in environmental assessment, where distributions of several parameters are conditioned on each other in a hierarchical fashion.  In the analysis of mercury contamination in king mackerel, this made it possible to obtain contaminant concentrations more precisely than with marginal distributions that ignore statistical interdependency between model parameters~\cite{voit95}.

Monte Carlo simulations of a quite different nature can be found in the  numerical application of mathematically controlled comparisons~\cite{alves00a,alves00b,alves00c}.  Mathematically controlled comparison~\cite{savageau72,irvine91,hlavacek98} is a technique to study in quantitative terms models of complex biological networks with alternative designs.  Well-worked applications of this technique led to the discovery of design principles for biosynthetic pathways~\cite{savageau72}, gene networks~\cite{savageau74,savageau01}, and immune networks~\cite{irvine85a,irvine85b}.  Mathematically, the method is based on S-systems within the power-law formalism~\cite{savageau76}, which leads to models of alternative designs that are often amenable to analytical solution and general conclusions.  The numerical extension is motivated by the need to eliminate some uncertainties associated with the classical analytical approach (e.g., numerical comparison of alternatives that depends on specific parameter values), and also by the need to address more complicated situations when power-law models are intractable analytically such as models of elementary signal transduction modules based on covalent modification of proteins.  In these cases, the idea is to sample parameters of such models from their (generally unknown) distributions and to evaluate statistical properties of model output properties, like steady-state levels of variables and logarithmic gains.  Although parameters were sampled from uniform distributions in recent applications of numerical mathematically controlled comparisons~\cite{alves00b,alves00c,alves00d,alves01}, sampling parameters from suitably chosen {\em S} distributions will be more appropriate in situations when the distributions are clearly not uniform.

Another application of the extroadinary flexibility of the {\em S} distribution in modeling various data structures is the study of distributional trends, which allows one to make inferences about the dynamics of probabilistic models of some stochastic processes.  Such models often include both stochastic and deterministic components~\cite{voit96}.  An example of such an approach is the analysis of trends in the distributions of tree sizes with their age~\cite{voit00b}.  Observations show that the distribution of tree trunk diameters changes with their age, even as radically as reversing the skewness.  By combining a deterministic component of the process (growth function of a tree) with a stochastic one (distribution of tree trunk diameter in the population) these authors were able to predict the change in distributional shape as a function of age, which is in good agreement with the observed data.  Along the same lines, \citeasnoun{sorribas00} considered growth trends in children (e.g. weight) with the modification that the trends of distributional parameters were established by regression, rather than from a deterministic model.  Similar technique also was used in~\cite{voit95,voit96a}.

\section{Discussion}

In this article we addressed an important issue in statistical modeling, that of statistical inference about a distribution.   The {\em S} distribution, which has been demonstrated to be a highly useful tool for various kinds of statistcal modeling, so far has lacked an estimator that is relatively straightforward, that makes a minimum reduction of information in the data, and that is amenable for goodness-of-fit analyses.  The MD estimators that we propose fill this gap.  Several features of the MD estimators make them viable alternatives to other methods.  Under some regularity conditions, MD estimators are strongly consistent~\cite{sahler70}.  This result applies in particular to the supremum and quadratic distance functions considered in this article, and to the {\em S} distribution which has a continuous d.f.  Also, MD estimators are invariant with respect to transformation of the estimand, a property that they share with ML estimators.  Finally, the following feature makes MD estimators especially relevant in the context of the {\em S} distribution.  When the hypothesized model does not belong to the class of parameterized d.f. models that generate the sample (i.e., when the model is wrong), MD estimators provide the ``best approximation'' from the class of models~(\ref{fclass}).  This is fully in the spirit of the {\em S} distribution being the best approximating distribution of the unknown population.  This feature is not shared by other estimation methods, including ML and the method of moments~\cite{parr81}.  Also, MD estimators are natural candidates for use in goodness-of-fit tests if the distance function is used as a goodness-of-fit statistic, as noted by \citeasnoun{bolthausen77}.

Given the numerical nature of the {\em S} distribution and the estimation method, large-scale Monte Carlo siumulations will be needed to establish properties of the estimator and the power of different distance metrics relative to other estimation methods, maximum likelihood in particular.  However, the example we have given of inference from a random sample generated from an {\em S} distribution shows the utility of the new estimator.   First, BCa bootstrap confidence intervals seem to require many more resamples than is thought appropriate for a general case: even the 4000 resamples reported here are clearly not enough to reduce the variance of estimates of the confidence intervals endpoints for all parameters.  Variance estimated with jackknife-after-bootstrap is even more erratic for quadratic than for supremum statistics.  Second, bootstrap distributions of MD estimates are highly skewed for all functions except Kolmogorov-Smirnov, which makes it the only one to have confidence intervals of reasonable length and shape and to be overall more trustworthy.  This observation goes along with the finding that consonance sets for location and scale parameters based on the Kolmogorov-Smirnov statistic have some desirable properties, i.e. they are finite and convex~\cite{salvia80}.  Reasons for the rather erratic behavior of the quadratic distance functions seem related to the fact that they are generally more sensitive to deviations from the model than supremum functionals.  During bootstrapping, variability of resamples is more pronounced in the tails of the bootstrap distributions and that is where the estimation of confidence interval endpoints (as sample quantiles) takes place.  Thus the variability of these estimates should be greater with the quadratic functionals.  For these reasons, the quadratic functionals in the context of {\em S} distributions seem to be less robust for estimation purposes, in contrast to what was found for more traditional location-scale families in Monte Carlo studies~\cite{parr80}.  Finally, expensive BCa intervals can be replaced by at least ten-fold less expensive extreme percentiles intervals, with little loss of accuracy.

We have shown that MD estimation coupled with bootstrap analysis of goodness-of-fit makes the {\em S} distribution a valuable tool for various kinds of Monte Carlo statistical modeling.

\section*{Acknowledgements}

This work was supported in part by U.S. Public Health Service Grant RO1-GM30054 from the National Institutes of Health.

%\newpage
%\pagestyle{empty}

%\section*{}

%\begin{figure}
%\caption{Empirical (dots) and theoretical {\em S} distribution (lines) d.f.s calculated with the MD estimates.  See Table~\ref{pointexample} for population and estimated parameters.} \label{fig1}
%\end{figure}

\section*{}

%\linespread{1}

\begin{sidewaystable}
\begin{tabular}{lllllll}
\hline
Parameter& Population& Combined estimate& \multicolumn{4}{l}{Minimum distance estimate}\\
&&& KS& KP& CVM& Wat \\
\hline
$\hat{g}$& 0.5& 0.810135& 0.516187& 0.504587& 0.628322& 0.633029\\
$\hat{h}$& 1.6& 1.40772& 1.49055& 1.36926& 1.57269& 1.62646\\
$\hat{\alpha}$& 1.0& 0.923027& 1.18409& 1.25795& 1.29237& 1.25799\\
$\hat{x}_0$& 0.0& 0.0212869& -0.00788733& 0.0313802& 0.0226832& 0.0117285\\
$D$& 0.0565035& 0.233542& 0.0398301& & \\
$V$& 0.0939742& 0.410878& & 0.0790387& \\
$W^2$& 0.0338124& 1.76827& & & 0.0197829& \\
$U^2$& 0.0324051& 1.74806& & & & 0.0197229 \\
\hline
\end{tabular}
\caption{Parameter estimates from an {\em S} distribution sample data ($n=100$).  Estimates obtained by a combination of existing methods (Combined estimate) and by minimizing KS (Kolmogorov-Smirnov), Kuiper (KP), Cram\'{e}r-von Mises (CVM) or Watson (Wat) distance metrics (Minimum distance estimate).  See text for details of calculation.} \label{pointexample}
\end{sidewaystable}

\section*{}

\begin{sidewaystable}
\begin{tabular}{lllll}
\hline
Parameter& \multicolumn{4}{l}{Confidence interval} \\
& KS& KP& CVM& Wat \\
\hline
$g$& (0.467339, 0.818564)${}^\ast$& (0.40623, 0.952086)& (0.179366, 0.96643)& (0.0777369, 1.01187)\\
 & (0.271795, 0.846612)${}^\dagger$& (0.289727, 0.935185)& (0.208938, 0.988865)& (0.203831, 1.15758)\\
$h$& (1.329, 2.17525)& (1.2042, 2.4769)& (0.943656, 7.29396)& (0.869806, 7.98307) \\
 & (0.975641, 2.01904)& (1.13551, 2.1405)& (0.814594, 3.64494)& (0.869556, 22.2877)\\
$\alpha$& (0.882613, 1.56611)& (0.650582, 1.65693)& (0.404169, 6.02715)& (0.398919, 20.7098) \\
 & (0.946866, 1.551)& (1.02877, 1.56108)& (0.503189, 13.1778)& (0.449198, 9.42905)\\
$x_0$& (-0.146193, 0.156423)& (-0.33061, 0.258266)& (-0.178406, 0.159036)& (-0.258809, 0.164895) \\
 & (-0.199951, 0.261195)& (-0.27011, 0.250137)& (-0.157073, 0.262181)& (-0.235997, 0.271462) \\
$D$& (0.0361164, 0.0718024)& & & \\
& $p=0.846154$& & & \\
 & (0.0241676, 0.0632652)& & & \\
 & $p=0.641026$& & & \\
$V$& & (0.0668382, 0.130067)& & \\
&& $p=0.871795$& & \\
 && (0.0533178, 0.108985)& & \\
 && $p=0.512821$& & \\
$W^2$ &&& (0.0108397, 0.0625526)& \\
&&& $p=0.615385$& \\
&&& (0.012497, 0.0689029)& \\
&&& $p=0.769231$& \\
$U^2$ &&&& (0.0120847, 0.0504255)\\
&&&& $p=0.666667$ \\
&&&& (0.0112161, 0.0971751)\\
&&&& $p=0.717949$ \\
\hline
\multicolumn{5}{l}{${}^\ast$Top entries calculated with nonparametric bootstrap}\\
\multicolumn{5}{l}{${}^b\dagger$Bottom entries calculated with parametric bootstrap}\\
\hline
\end{tabular}
\caption{Equitailed extreme-percentile bootstrap confidence intervals. Intended coverage of 95\% for minimum distance parameter estimates from sampled data ($n=100$) obtained from the {\em S} distribution with parameters $g=0.5$, $h=1.6$, $\alpha=1.0$, $x_0=0.0$.  Methods include KS (Kolmogorov-Smirnov), Kuiper (KP), Cram\'{e}r-von Mises (CVM) and Watson (Wat) goodness-of-fit statistics.  See text for details of calculation.} \label{extremeexample}
\end{sidewaystable}

\section*{}

\begin{sidewaystable}
\begin{tabular}{lllll}
\hline
Parameter& \multicolumn{4}{l}{Confidence interval} \\
& KS& KP& CVM& Wat \\
\hline
$g$& (0.296367, 0.681915)${}^\ast$& (0.204395, 0.693703)& (0.0257174, 0.888425)& (0.00721408, 0.955644)\\
 & (0.259294, 0.70718)${}^\dagger$& (0.274027, 0.846456)& (0.104614, 0.959833)& (0.0916488, 1.04173)\\
$h$& (1.13667, 1.79035)& (0.981214, 1.65894)& (1.01331, 6.36995)& (0.96578, 10.1783) \\
 & (1.14481, 1.94514)& (0.960117, 1.90602)& (0.915669, 5.68933)& (0.885967, 6.62075)\\
$\alpha$& (0.835117, 1.35433)& (0.914392, 1.63598)& (0.331079, 5.24991)& (0.314168, 4.48109) \\
 & (0.846121, 1.31254)& (0.908877, 1.63787)& (0.354572, 7.13847)& (0.353748, 7.24277)\\
$x_0$& (-0.196492, 0.151388)& (-0.188277, 0.280725)& (-0.133676, 0.196232)& (-0.173835, 0.181506) \\
 & (-0.216363, 0.225872)& (-0.182983, 0.386315)& (-0.190276, 0.251492)& (-0.202019, 0.278002) \\
$D$& (0.0282195, 0.0448809)& & & \\
& $p=0.8845$& & & \\
 & (0.02683, 0.0531105)& & & \\
 & $p=0.634$& & & \\
$V$& & (0.0511786, 0.0907438)& & \\
&& $p=0.86125$& & \\
 && (0.0572763, 0.111431)& & \\
 && $p=0.55325$& & \\
$W^2$ &&& (0.00993233, 0.0356645)& \\
&&& $p=0.64$& \\
&&& (0.00744214, 0.0311884)& \\
&&& $p=0.81$& \\
$U^2$ &&&& (0.00989892, 0.037251)\\
&&&& $p=0.59575$ \\
&&&& (0.00707178, 0.0322339)\\
&&&& $p=0.7805$ \\
\hline
\multicolumn{5}{l}{${}^\ast$Top entries calculated with nonparametric bootstrap}\\
\multicolumn{5}{l}{${}^\dagger$Bottom entries calculated with parametric bootstrap}\\
\hline
\end{tabular}
\caption{Equitailed BCa percentile bootstrap confidence intervals.  Intended coverage of 95\% for minimum distance parameter estimates from sampled data ($n=100$) obtained from the {\em S} distribution with parameters $g=0.5$, $h=1.6$, $\alpha=1.0$, $x_0=0.0$.  Methods include KS (Kolmogorov-Smirnov), Kuiper (KP), Cram\'{e}r-von Mises (CVM) and Watson (Wat) goodness-of-fit statistics.  See text for details of calculation.} \label{bcaexample}
\end{sidewaystable}

\section*{}

\newpage

\end{document}